# PlomBOX - development of a low-cost CMOS device for environmental monitoring

**Aguilar-Arevalo A.[7], Alba Posse E.[1,10], Alvarez M.[1], Arnaldi H.[2], Asorey H.[2,8], Bertou X.[2], Colque A.[1], Deisting A.[6], Dias A.[6,*], D'Olivo J.C.[7], Favela-Pérez F.[7], Gándola Y.[1], Garcés E. A.[4], Gasulla J.[1,9], Gómez Berisso M.[2], González Muñoz A.[4], Guerra-Pulido J. O.[7], Gutierrez S.[2], Jois S.[6], Lipovetzky J.[2], Lovera J.[2], Lovino M.B.[2], Marín-Lámbarri D.J.[4], Marpegan L.[2], Martín D.[1], Martinez Montero M.[7], Mejía Muñoz S.[1], Monroe J.[6], Nadra A.[1,10], Paling S.[5], Pregliasco R.[2], Rumi G.[2], Rossen A.[3], Santos J.[1], Scovell P. R.[5], Tallis M.[2], Teijeiro A.[1], Triana M.[1], Vázquez-Jáuregui E.[4]**

[1] Universidad de Buenos Aires, Facultad de Ciencias Exactas y Naturales, Instituto de Biociencias, Biotecnología y Biología Traslacional (iB3), Intendente Güiraldes 2160, Ciudad Universitaria, C1428EGA, Buenos Aires, Argentina.

[2] Centro Atómico Bariloche and Instituto Balseiro, Comisión Nacional de Energía Atómica (CNEA), Consejo Nacional de Investigaciones Científicas y Técnicas (CONICET), Universidad Nacional de Cuyo (UNCUYO), San Carlos de Bariloche, Argentina

[3] Laboratorio de Experimental de Tecnologías Sustentables. Centro de Tecnología del Uso del Agua. Instituto Nacional del Agua. Au Ezeiza-Cañuelas km 1.6 CP (1804) Pcia de Buenos Aires.

[4] Instituto de Física, Universidad Nacional Autónoma de México, A. P. 20-364, México D. F. 01000, Mexico

[5] Boulby Underground Laboratory. Science and Technology Facilities Council, England, United Kingdom.

[6] Royal Holloway. University of London, United Kingdom.

[7] Instituto de Ciencias Nucleares, Universidad Nacional Autónoma de México, CDMX, México

[8] Instituto de Tecnologías en Detección y Astropartículas, Comisión Nacional de Energía Atómica (CNEA), Consejo Nacional de Investigaciones Científicas y Técnicas (CONICET), Universidad Nacional de San Martín (UNSAM), Buenos Aires, Argentina.

[9] Centro de Investigaciones del Medio Ambiente (UNLP-CONICET), La Plata, Buenos Aires, Argentina

[10] Consejo Nacional de Investigaciones Científicas y Técnicas. Godoy Cruz 2290 C1425FQB, Buenos Aires, Argentina.

*corresponding author: Adriana Dias
e-mail: adriana.dias.2011@live.rhul.ac.uk

**Abstract**

This paper reports on the development of a novel CMOS device employing lead-sensing bacteria to assay lead in drinking water. The objective of the PlomBOX project is to develop a low-cost sensor (£10) which can expedite access to on-demand assay methods and thus help mitigate lead intake through contaminated drinking water. The project follows three development paths: a) Certain bacteria can fluoresce or change colour when in the presence of lead. A genetically modified strain of *Escherichia coli* sensitive to lead concentrations up to 10 ppb is being developed. This constitutes the biosensor that fluoresces in proportion to the presence of lead. b) Bacteria response is imaged using a microprocessor (ESP32) with a camera module. This constitutes the optical metrology component of the PlomBOX. c) Data acquisition and control of the PlomBOX is achieved through a Bluetooth connection with the PlomApp, a custom-developed mobile phone application. Data are sent from the PlomApp to a database where a bespoke automated analysis software provides a result of the lead concentration in a sample of water. This paper reports on the instrumentation challenges of developing the electronics for the PlomBOX and on the first prototype.

## 1. Introduction

Lead is one of ten chemicals that the World Health Organisation (WHO) has identified as a major public health concern [1]. Lead poisoning affects multiple systems in the human body and is particularly harmful in young children [1]. The WHO has set an upper limit of 0.01 mg/l (10ppb) in drinking water [2], although there is no level of lead exposure that is known to be harmless [1].

In low and middle-income countries, an estimated 26 million people are at risk from lead exposure in drinking water [3]. In developed countries lead screening tests are carried out in both public and private water supplies [4]. In low- and middle-income countries access to accredited laboratories is more limited and may be prohibitively expensive.

Biosensors have been shown to successfully indicate concentrations of arsenic in drinking water at the 10ppb scale [5]. Detecting such small indicators of contamination is a challenge that is common in dark matter (DM) experiments, which must assay detector components for trace radioactivity at a level at or below 1 ppb of lead, in the presence of a background of natural radioactivity. This project combines these two approaches with the aim of developing a low-cost sensor for lead in drinking water. A low-cost sensor would expedite access to on-demand assay methods and could help mitigate lead intake through contaminated drinking water. Low-cost sensors are widely available, e.g. complementary metal-oxide-semiconductor (CMOS) sensors in mobile phone cameras. Furthermore, low-cost microprocessors with CMOS cameras are available off-the shelf (COTS) and can be used by non-experts in the field.

The principle of the PlomBOX operation is that the fluorescence/colour change of a specifically engineered bacteria in contact with a water sample will correspond to the sample's lead level. This can be measured precisely via fluorescence/colour spectroscopy using a CMOS sensor – Figure 1.

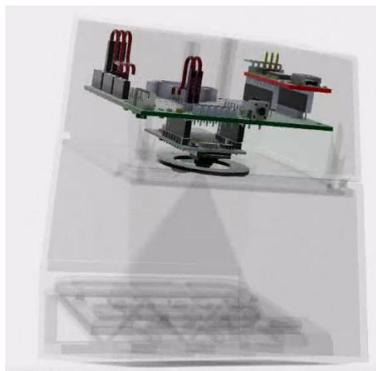

**Figure 1** CAD model of the PlomBOX prototype with the CMOS sensor measuring the fluorescence of a bacteria strain.

This paper will discuss: selection of an optimal bacteria that senses lead in water; development of a light-sensitive device to detect the enhanced fluorescence signal or colour change in the water sample; and data acquisition from the measurement device using a mobile phone.

## 2. Methodology

### 2.1. Design of lead sensitive bacterium

To develop a lead sensitive bacterium, a genetically modified strain of *E. coli* was generated. It was chosen due to its wide use in laboratories, fast growth, and availability of molecular tools to modify its physiology.

Three genetic constructions responsive to lead were designed. They have a regulatory zone that includes the PbrR repressor as a lead-specific binding protein and the promoter *pbr* from the lead resistance operon of *Cupriavidus metallidurans* CH34 with a downstream reporter. Upon PbrR repressor binding to lead, the reporter gene is expressed, and this generates a signal. The type of signal depends on the reporter: *β-galactosidase,* colorimetric; Green Fluorescent Protein (sfGFP), fluorescent; and NanoLuciferase (Nluc), luminescence. Three different strains of *E. coli* were evaluated: DH5alpha, Top10 and BL21.

For the lead sensitivity measurement, the bacteria were grown overnight in a Luria–Bertani (LB) medium and normalized to an optical density at 600 nm (OD600) = 1.5. An inoculum of 20µl of this culture was diluted in 800µl of final volume with solutions containing culture medium and lead until reaching a concentration of 0, 10 and 100 ppb for induction. The measurement was carried out at 37°C in triplicate, in a 96-well plate and a Tecan Infinite® 200 PRO plate reader. Fluorescence for sfGFP is measured at an $\lambda_{ex}$ (excitation wavelength) of 485nm and an $\lambda_{em}$ (emission wavelength) of 535nm. The blue colour is measured by OD at 620nm. For the chromogenic method, generation of blue colour is measured by adding the X-gal reagent (the blue colour is directly related to the amount of *β-galactosidase* reporter protein produced during the assay). For the fluorogenic method, the fluorescence intensity of the sfGFP is measured. For the light emitting method, the goal was to measure the light produced by the Nluc protein, although the method has not yet been evaluated. In addition to reporter measurements, the bacteria growth is monitored by OD at 600nm (OD600).

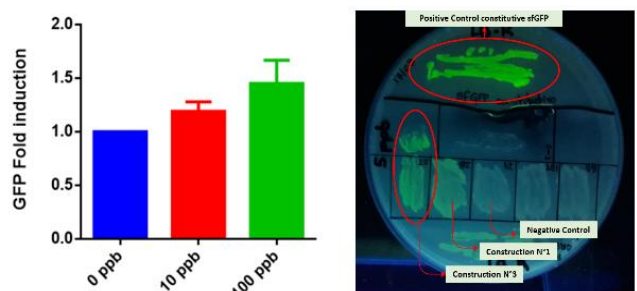

**Figure 2** Measurement of GFP protein lead dose response. Left: GFP Fold Induction (bacteria concentration increase %) vs Lead concentration in ppb for Construction n3. Right: Petri dish with several *E. coli* colonies expressing GFP. Negative Control and Positive control correspond to an untransformed *E. coli* strain and a GFP overexpressing strain, respectively.

Measurements using these three reporters return high variability between data sets. This had previously been observed in [5] and is typical of biological systems. The obtained response to lead is very small and therefore the variability of the system ends up masking this response. To mitigate the observed high variability between datasets, two new genetic constructs were developed. Both constructions use pbrR2 protein, homologous to pbrR. One has sfGFP as the reporter protein and was kept similar to the wild gene type. The other construct contains several genetic modifications thought to improve lead sensitivity response, as well as the fluorescent mCherry protein to be used as an internal growth control. These new genetic constructs were measured using the same method described previously. mCherry fluorescence was monitored at $\lambda_{ex}$ 570nm and $\lambda_{em}$ at 600nm.

All these constructions show that a measurement of lead concentration can be made in the WHO relevant concentration range by a colour / fluorescence change of genetically modified bacteria - Figure 2. However, results demonstrate small change in reporter expression relative to lead concentration changes. An attempt to reduce experimental variability is being made with the goal of obtaining more robust measurements.

*2.2 Development of optical metrology components*

The development of the optical metrology components began with the characterisation of COTS detectors with As-sensitive bacteria used in [5], and development of a simplified optical measurement system.

Illumination conditions are an important aspect of the bacteria light emission measurement. For fluorescent reporters, it is important to control $\lambda_{ex}$ within a narrow band from 10 to 30 nm, centred in the required $\lambda_{ex}$. For colorimeter reporters, illumination is made with a broad-spectrum source. Interference filters are also required for fluorescent and colorimeter reporters. With these considerations, the spectral characteristics of commercial light emission devices and detection components (colourimeters, CMOS image sensors, and photodiodes with and without and filters) were measured. [5] quantified an emission at 515 nm wavelength of these bacteria excited with a wavelength of 475 nm. Measurements were developed of a ratio of intensities from the Red, Green, Blue (RGB) pixels of a CMOS with an optimized incident excitation wavelength spectrum. Three different measurement setups were operated in relation to the initial reporter strains described in Section 2.1: a) a single-pixel colourimeter sensor, intended to determine the *β-galactosidase* concentration evolution, measured as intensity of blue colour change as a function of time; b) a CMOS image sensor, to obtain precise colourimetric sampling by simultaneous measurement at different RGB pixels, and also to measure the GFP reporter response; and c) a photometric sensor based on a Multi-pixel photon counter, or silicon photomultiplier (MPPC/SiPM).

These studies permitted the development of a simulation to understand the response of optical metrology components to the mCherry reporter. The simulation accounted for the fact that the absorption spectra of the protein serve as input to calculate the fraction of the incident light coming from the LED that gets absorbed at each wavelength, given a protein concentration, using the Beer-Lambert law. Its integral over all wavelengths gives the probability that a photon coming from the LED is absorbed. This quantity, together with the quantum yield of the protein, modulates the amplitude of the resulting fluorescence spectrum. Figure 3 describes the two contributing factors towards the signal that is received by the sensor: light emitted by the bacteria fluorescence and light emitted by the LED. This study showed that it was important for the optical sensor to have high efficiency in the range of 500 to 600nm.

Following the bacteria strain development and optical metrology research of different devices, the electronic system was assembled for the prototype. The device is composed of a custom-made Printed Circuit Board (PCB) and a commercial board, the ESP32-CAM. The custom-made board provides power and the connection interfaces for firmware updates to the ESP32-CAM device, and it contains the different sensors used to complete the data capture. These sensors include a DS18B20 device for the temperature sensing and PL9823 programmable LEDs, for the sample illumination. Additionally, the custom-made board provides interconnection interfaces for adding new sensors through standard protocols (1-Wire and I2C). The ESP32-CAM module was chosen because it is a powerful yet inexpensive microcontroller with advanced features which include Bluetooth, Wi-Fi, and multipurpose General-purpose input/output (GPIO) ports. The ESP32-CAM device also contains an integrated video camera, the OV2640 module with a 2-megapixel sensor and microSD card socket.

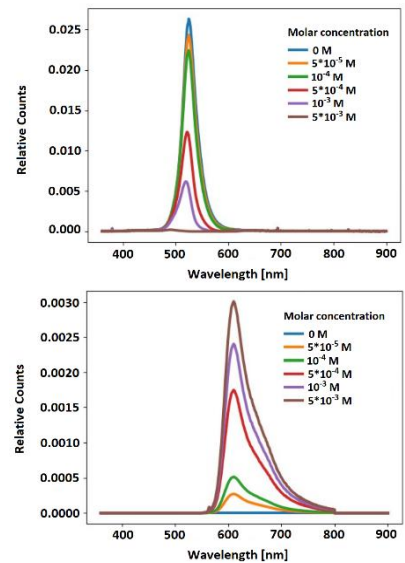

**Figure 3** Top: Efficiency (%) vs Wavelength (nm) of an LED emitted photon. Bottom: Efficiency (%) vs Wavelength (nm) of emitted light from mCherry fluorescence. In both plots the curves represent the molar concentration of bacteria.

*2.3 Development of data acquisition and analysis tools*

To acquire data from the PlomBOX, a mobile phone application, PlomApp, was developed in Android Studio using Java. The application controls the prototype through the PlomBOX firmware running on the ESP32, and acquires images from the PlomBOX via Bluetooth, while collecting meta-data from the smartphone geo-location sensors. The PlomApp begins by sending a command via Bluetooth to the ESP32-CAM to acquire data. Prior to sending the data to the phone, the images are converted to JSON strings on the ESP32-CAM. The data are sent from the ESP32-CAM to the PlomApp via Bluetooth. Once received on the app, the data is sent to a remote data server via Message Queuing Telemetry Transport (MQTT) protocol. A secure communication protocol was developed relying on TLS certificates for individual users. As this solution is not easily scalable within the Android application, different alternatives such as Android Firebase user registration services are being considered.

In the data server, analysis tools developed using Python process the incoming data and return Hue, Value and Saturation (HSV) values as a result. Previous tests in [5] showed that qualitative image analysis, performed by visual inspection, was possible. This suggested the use of the HSV colour base since it closely resembles the human perceptual experience and at the same time it is quantifiable. Two of the three parameters of the colour base show small variation (H and V) at constant lightning conditions and camera settings, while the saturation (S) increases as the lead concentration increases – Figure 4. This provides a zeroth order approximation for a single parameter that can quantify lead concentration. This approximation is still being developed. After the HSV analysis processes incoming data, HSV results for each image are returned to the phone.

The first prototype conceptual test allowed the integration of the development paths described in sections 2.2 and 2.3. The electronic board of the PlomBOX device was assembled and aligned in its 3D-printed light-tight case. The ESP32 CAM was paired successfully through Bluetooth with the PlomApp, with the latter successfully controlling the camera. Additionally, telemetry and colour images acquisition were tested, using the onboard ESP32 camera and sensors. The acquired data was encoded in the JSON format and stored locally before transferring it to the PlomApp through Bluetooth. The app also succeeded in transferring the JSON packages from the phone to the PlomBOX server using the MQTT secured protocol. Additionally, the PlomApp decoded and displayed the acquired images on the phone's screen – Figure 5. Finally, the server received and stored the images sent through the PlomApp and ran a quick HSV analysis of the data, whose results were sent back to the app.

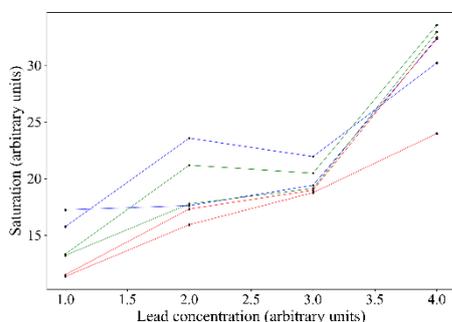

**Figure 4** Saturation vs Lead concentration of a colorimetric bacteria sample measurement from [5], in arbitrary units. Each line represents one bacteria culture in the presence of lead, with varying pre-culture conditions.

### 3. Outlook and conclusion

This paper reports and demonstrates the technical feasibility of the PlomBOX concept. The different bacteria constructions show that a measurement of lead concentration can be made in the WHO relevant concentration range by a colour / fluorescence change of genetically modified bacteria. Additionally, a low-cost electronic metrology system using a ESP32-CAM module has been developed which is inexpensive yet powerful. Finally, a mobile phone application, PlomApp, is being developed to undertake all data acquisition and slow control functions of the PlomBOX, whilst requiring minimum user input. Once the design is finalized, we plan a serial production of 10-20 units at a unit cost of ~£10. Additionally, the project aims to undertake a field test, to optimize user experience and utility by obtaining a field evaluation and survey on user experience. Furthermore, we will cross-calibrate the field test results with external lead assay from chemical water analysis as well as techniques used for radioassay in dark matter experiments, which include the use of Germanium and Inductively coupled plasma mass spectrometry (ICP-MS) detectors.

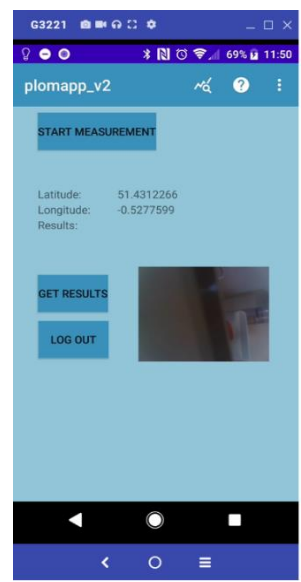

**Figure 5** PlomApp screenshot showing image retrieved from ESP32 CAM.

### 4. Acknowledgements

This research was funded by the GCRF Translation Award EP/T015586/1, STFC Global Challenges Research Fund (Foundation Awards, Grant ST/R002908/1) and by DGAPA UNAM grants PAPIIT-IT100420 and PAPIIT-IN108020, and by CONACyT grants CB-240666 and A1-S-8960.